# Tunable Spin-Orbit Torques in Cu-Ta Binary Alloy Heterostructures

Tian-Yue Chen, Chun-Te Wu, Hung-Wei Yen, and Chi-Feng Pai[*]

*Department of Materials Science and Engineering, National Taiwan University, Taipei 10617, Taiwan*

The spin Hall effect (SHE) is found to be strong in heavy transition metals (HM), such as Ta and W, in their amorphous and/or high resistivity form. In this work, we show that by employing a Cu-Ta binary alloy as buffer layer in an amorphous $Cu_{100-x}Ta_x$-based magnetic heterostructure with perpendicular magnetic anisotropy (PMA), the SHE-induced damping-like spin-orbit torque (DL-SOT) efficiency $|\xi_{DL}|$ can be linearly tuned by adjusting the buffer layer resistivity. Current-induced SOT switching can also be achieved in these $Cu_{100-x}Ta_x$-based magnetic heterostructures, and we find the switching behavior better explained by a SOT-assisted domain wall propagation picture. Through systematic studies on $Cu_{100-x}Ta_x$-based samples with various compositions, we determine the lower bound of spin Hall conductivity $|\sigma_{SH}| \approx 2.02 \times 10^4 \left[\hbar/2e\right] \Omega^{-1} \cdot m^{-1}$ in the Ta-rich regime. Based on the idea of resistivity tuning, we further demonstrate that $|\xi_{DL}|$ can be enhanced from 0.087 for pure Ta to 0.152 by employing a resistive TaN buffer layer.

[*] Email: cfpai@ntu.edu.tw

The spin Hall effects (SHE) [1-3] in heavy transition metals (HM) are known to be strong enough to generate sizable pure spin currents from charge currents for SHE-induced spin-orbit torque (SOT) switching [4,5], magnetic oscillations [6,7], as well as chiral domain wall motion [8,9] on the adjacent ferromagnetic (FM) layers. The phenomenology of the SHE-generated transverse spin current (density) $J_s$ in a HM/FM heterostructure can be described by $J_s = (\hbar/2e) T_{int}^{HM|FM} \theta_{SH}^{HM} J_e^{HM}$, where $\theta_{SH}^{HM}$ and $J_e^{HM}$ represent the spin Hall ratio of HM layer and the longitudinal charge current density therein, respectively. $T_{int}^{HM|FM}$ is the spin transparency at the HM|FM interface [10], which takes any possible spin backflow [11] and/or spin memory loss [12] at the interface into account ($T_{int}^{HM|FM} = 1$ for the 100% spin transmission case). The resulting damping-like (DL) SOT efficiency acting upon the FM layer therefore can be expressed as $\xi_{DL} \equiv (2e/\hbar) J_s / J_e^{HM} = T_{int}^{HM|FM} \theta_{SH}^{HM}$ [13]. Among all HMs, 5d transition metals such as Pt [14], β-Ta [5], and β-W [15] related magnetic heterostructures are found to have giant DL torque efficiencies ($|\xi_{DL}| \sim 0.10 - 0.30$) due to their strong intrinsic spin-orbit interactions. It has also been shown that 5d HM-related alloys and oxides can generate sizable SHE and even possess greater spin Hall ratios than pure HMs. For instance, Hf(Al)-doped Pt [16], Au-doped Pt [17], PtMn [18], W-doped Au [19], and $WO_x$ [20] all show larger spin Hall ratios or $|\xi_{DL}|$ while compare to their pure HM counterparts. Even by oxidizing or doping dopants into light transition metals such as Cu, the spin Hall ratio can be significantly enhanced [21,22]. Apparently, engineering spin-Hall source materials by alloying or oxidation can possibly give rise to a more efficient SOT control over the adjacent FM layers.

In this work, we show that it is possible to linearly tune the DL-SOT efficiency of a magnetic heterostructure by controlling the buffer layer (spin Hall material layer) resistivity. The spin Hall material that we employ is a Cu-Ta binary alloy system, in which the spin Hall ratio can be adjusted by changing the relative $Cu_{100-x}Ta_x$ compositions. In Ta-rich regime ($x \geq 84$), the alloy becomes amorphous and serves as a suitable HM buffer layer for generating perpendicular magnetic anisotropy (PMA) for the subsequently deposited FM layer. We first demonstrate current-induced SOT switching in these perpendicularly-magnetized $Cu_{100-x}Ta_x$/FM heterostructures to verify the existence of giant SHE. Next, we systematically characterize the DL-SOT efficiency $|\xi_{DL}|$ of these $Cu_{100-x}Ta_x$/FM heterostructures using the hysteresis loop shift measurements [23]. $|\xi_{DL}|$ is found to be linearly proportional to the $Cu_{100-x}Ta_x$ alloy resistivity, which indicates an intrinsic mechanism and/or side-jump origin of the SHE, with a nearly constant spin Hall conductivity of $|\sigma_{SH}| \approx 2.02 \times 10^4 [\hbar/2e] \, \Omega^{-1} \cdot m^{-1}$ (lower bound, assuming $T_{int}^{HM|FM} = 1$). The critical SOT switching current density $J_c$ of these $Cu_{100-x}Ta_x$-based magnetic heterostructures can further be estimated from the characterized $|\xi_{DL}|$ and coercive fields $H_c$, which confirms the domain nucleation/propagation nature of SOT switching in micron-sized PMA devices. Lastly, to demonstrate the benefits of resistivity tuning, we engineer (increase) the resistivity of Ta buffer layer by nitrogen doping. The maximum DL-SOT efficiency of TaN is found to be $|\xi_{DL}| = 0.152 \pm 0.006$, while that of Ta control sample is $|\xi_{DL}| = 0.087 \pm 0.004$.

We prepared our magnetic heterostructures in a high vacuum sputtering chamber with base pressure of $3 \times 10^{-8}$ Torr. DC and RF magnetron sputtering with 3 mTorr of Ar working pressure were employed for depositions of metallic and oxide (MgO) layers, respectively. The binary alloy buffer layer $Cu_{100-x}Ta_x$ was deposited by confocal co-sputtering from a pure Cu and a pure Ta target. The relative compositions can be tuned by the relative sputtering powers of two targets. Multilayer stack heterostructures $Cu_{100-x}Ta_x(4)/W(0.5)/Co_{20}Fe_{60}B_{20}(1.4)/Hf(0.5)/MgO(2)/Ta(2)$ (numbers in the parenthesis are in nm) were deposited onto thermally-oxidized silicon substrates, with $x$ ranges from 20 to 90, as shown in Fig. 1(a). We annealed all our samples at 300 °C for one hour in vacuum to induce PMA of the CoFeB layer. The CoFeB(1.4) layer was sandwiched between a W(0.5) and a Hf(0.5) dusting layer since we found this combination beneficial for obtaining stable PMA for most cases, similar to previous reports [24]. The resistivity of the co-sputtered $Cu_{100-x}Ta_x(4)$ layer was characterized by four-point measurements on unpatterned films. As shown in Fig. 1(b), the resistivity of $Cu_{100-x}Ta_x$ increases monotonically from 40 μΩ-cm to 240 μΩ-cm as we increase the Ta component from 10 at% to 90 at%, which is similar to the trend reported in previous studies [25]. There is also a slope change in the resistivity-to-composition plot at $x \simeq 70$, which indicates a possible transition from poly-crystalline phase to amorphous phase while the alloy is becoming Ta-rich.

The phase transition of $Cu_{100-x}Ta_x$ suggested by the resistivity trend was further verified by the cross-sectional images from high-resolution transmission electron microscopy (HR-TEM), as shown in Fig. 2(a) and (b). In the Cu-rich ($Cu_{70}Ta_{30}$) buffer layer, lattice fringes were largely observed but some domains are amorphous, as evidenced by the diffractograms in Fig. 2(a). Hence, the Cu-rich ($Cu_{70}Ta_{30}$) buffer layer is mainly polycrystalline with partially-mixed amorphous domains. It should be noted that the spacings of lattice fringes all correspond to the plane spacings of Cu. In contrast, the Ta-rich ($Cu_{10}Ta_{90}$) buffer layer has only an amorphous phase as shown in Fig. 2(b). The line-scan profiles (not shown) of the heterostructures also indicate that the Hf and W dusting layers are strongly intermixed with the CoFeB layer, therefore we label the W(0.5)/CoFeB(1.4)/Hf(0.5) layer as an effective ferromagnetic layer "FM" in the TEM images. The microstructure of the $Cu_{100-x}Ta_x$ buffer layer significantly affects the resulting magnetic anisotropy of the FM layer above. As presented in Fig. 2(c) and (d), the magneto-optical Kerr effect (MOKE) data clearly show that $Cu_{70}Ta_{30}$/FM/MgO is in-plane magnetized while $Cu_{10}Ta_{90}$/FM/MgO has PMA. The amorphous nature of the Ta-rich buffer layer makes the $Cu_{10}Ta_{90}$/FM interface smoother than that of the Cu-rich (poly-crystalline) heterostructure, which is beneficial for obtaining PMA. However, note that for both heterostructures shown here, the MgO layer is not crystalline, which suggests that the crystallinity of oxide layer is not a necessary condition for getting PMA [26].

Since most of the recent SOT switching studies have focused on magnetic heterostructures with PMA [4,5,27], we now turn our focus on $Cu_{100-x}Ta_x$/FM/MgO heterostructures in the Ta-rich regime ( $84 \leq x \leq 100$ ). We prepared Hall-bar samples of $Cu_{100-x}Ta_x$/FM/MgO with lateral dimensions of $5\mu m \times 60\mu m$ by photolithography and lift-off process. To demonstrate current-induced SOT switching in these heterostructures, as shown in Fig. 3(a), we first applied an in-plane field $H_x = 100\,\text{Oe}$ to either break the inversion symmetry (single domain picture [28]) or re-align the chiral domain wall moment (multi-domain picture [29]) to facilitate magnetic domain expansion. Then a DC charge current $I_{DC}$ was sent into the current channel of the Hall-bar device to generate a spin current from the SHE of $Cu_{100-x}Ta_x$ buffer layer. The resulting SHE-induced SOT acting upon the adjacent FM layer will switch the magnetization when $I_{DC}$ reaches a critical value. All samples with PMA ( $84 \leq x \leq 100$ ) can be reversibly switched by sweeping $I_{DC}$ up to $\pm 5\,\text{mA}$, and we found critical switching current of all samples to be $\sim \pm 2\,\text{mA}$ ( $J_c \sim \pm 0.8 \times 10^{11}\,\text{A} \cdot \text{m}^{-2}$ ). This suggests that the SHE from the $Cu_{100-x}Ta_x$ buffer layer is comparable to that from other pure HMs such as Pt, Ta, and W. A representative switching curve from a $Cu_6Ta_{94}$ device is shown in Fig. 3(b). The steps observed during current-induced switching can be considered as the evidence of domain wall motion, therefore we believe that the DL-SOT switching observed here is governed by domain nucleation and SOT-assisted domain (wall) expansion (propagation) [8,23,29].

The results of systematic studies on transport and magnetic properties of $Cu_{100-x}Ta_x$/FM/MgO heterostructures as functions of Ta content are summarized in Fig. 4. We first measured the resistance of Hall-bar devices to calculate the resistivity of buffer layer in its amorphous phase. The contribution from the effective FM layer has been subtracted. As shown in Fig. 4(a), the $Cu_{100-x}Ta_x$ buffer layer resistivity is linearly proportional to the Ta concentration, which is consistent with the unpatterned film result (the amorphous side of Fig. 1(b)). To systematically characterize the DL-SOT efficiency, we performed the hysteresis loop shift measurements [23,30,31] on $Cu_{100-x}Ta_x$/FM/MgO Hall-bar devices. In this type of measurement, the DL-SOT efficiency $\xi_{DL}$ can be estimated from the ratio of measured shift of out-of-plane hysteresis loop $H_{eff}^z$ to the applied current density $J_e$ by [23,32]

$$\xi_{DL} = \frac{2e}{\hbar} \mu_0 M_s \left(t_{FM} - t_{dead}\right) \cdot \left(\frac{2}{\pi}\right) \cdot \left(\frac{H_{eff}^z}{J_e}\right), \qquad (1)$$

where $M_s$, $t_{FM}$, and $t_{dead}$ represent saturation magnetization, nominal thickness, and dead layer thickness of the FM layer, respectively. These parameters are characterized by vibrating sample magnetometer (VSM) from unpatterned films to be $M_s \approx 1500\,\text{emu/cm}^3$ ($1.5\times10^6$ A/m in SI units) and $t_{FM} - t_{dead} \approx 0.7$ nm. The estimated magnitude of DL-SOT efficiency $|\xi_{DL}|$ is found to be linearly proportional to the Ta content, as shown in Fig. 4(b). Note that the sign of $\xi_{DL}$ is actually negative, consistent with the negative spin Hall ratio of Ta [5]. Since both the resistivity $\rho_{CuTa}$ and DL-SOT efficiency $|\xi_{DL}|$ are proportional to the Ta content of buffer layer, it is obvious that $|\xi_{DL}| \propto \rho_{CuTa}$ in our heterostructures.

According to the theory of SHE, if the phenomenon is driven by intrinsic mechanism (or side-jump scattering) [1,33] then the spin Hall ratio is related to the spin Hall conductivity $\sigma_{SH}$ and the conductivity $\sigma_{HM}$ (or its inverse, resistivity $\rho_{HM}$) of the HM layer by $\theta_{SH}^{HM} = \sigma_{SH}/\sigma_{HM} = \sigma_{SH} \cdot \rho_{HM}$. The resulting DL-SOT efficiency $|\xi_{DL}| = |T_{int}^{HM|FM} \theta_{SH}^{HM}| \propto |\theta_{SH}^{HM}| \propto \rho_{HM}$, provided that the spin transparency at the interface being unchanged. Therefore, our discovery of $|\xi_{DL}| \propto \rho_{CuTa}$ suggests that the dominating mechanism of the SHE in $Cu_{100-x}Ta_x$ buffer layer is intrinsic. By assuming a perfect spin transmission at the HM|FM interface, $T_{int}^{HM|FM} = 1$, the lower bound of the spin Hall conductivity is estimated to be $|\sigma_{SH}| \approx 2.02 \times 10^4 \, [\hbar/2e] \, \Omega^{-1} \cdot m^{-1}$ from the ratio between $|\xi_{DL}|$ and $\rho_{CuTa}$. If we further assume that the contribution to SHE is solely coming from Ta, then this estimated spin Hall conductivity of Ta is fairly close to the prediction from first-principle calculation [34]. It is also interesting to note that our discovery of a linear dependence between spin Hall ratio and buffer layer resistivity ($|\theta_{SH}^{HM}| \propto \rho_{HM}$) is different from the quadratic dependence ($|\theta_{SH}^{HM}| \propto \rho_{HM}^2$) reported by Hao and Xiao for Ta-based PMA heterostructures [35], in which the resistivity was tuned by temperature. However, $\left|\theta_{SH}^{HM}\right| \propto \rho_{HM}$ is not an uncommon trend for sputtered heavy transition metals in the moderately dirty (resistive) regime, as can be seen from the cases of Pt [13,36].

We summarize the (out-of-plane) coercive field $H_c$ and the (current-induced SOT) critical switching current density $J_c$ of all $Cu_{100-x}Ta_x$/FM/MgO samples with PMA in Fig. 4(c) and (d), respectively. Conventionally, $J_c$ of the current-induced SOT switching in a PMA heterostructure is considered to be related to the DL-SOT efficiency $\left|\xi_{DL}\right|$, the effective anisotropy field $H_k$, and the externally applied in-plane field $H_x$ through [37]

$$J_c = \frac{2e}{\hbar}\mu_0 M_s \left(t_{FM} - t_{dead}\right) \cdot \left(\frac{H_k}{2} - \frac{H_x}{\sqrt{2}}\right) / \left|\xi_{DL}\right|. \tag{2}$$

However, this expression is only suitable for FM layer being single domain such that the switching behavior can be described by a macro-spin model. In most experimental cases, the samples prepared are micron-sized and the magnetization switching process is dominated by domain nucleation and domain wall propagation [23,29,38]. Therefore, in our case, the critical switching current density should be expressed as

$$J_c = \frac{2e}{\hbar} \mu_0 M_s \left(t_{FM} - t_{dead}\right) \cdot \left(\frac{2}{\pi}\right) \cdot \left(\frac{H_c}{|\xi_{DL}|}\right), \tag{3}$$

in which the coercive field (or the depinning field) $H_c$ and the DL-SOT efficiency $|\xi_{DL}|$ are the two major factors in determining $J_c$. Using Eqn. (3), we estimated $J_c$ with the data in Fig. 4(b) and (c) and plot the results alongside with the experimentally determined $J_c$ in Fig. 4(d). Though $|\xi_{DL}|$ increases while the buffer layer becomes more Ta-rich, $H_c$ also has a similar growing trend. Therefore, the resulting $J_c$ is almost constant ($\sim 0.8 \times 10^{11} \text{A} \cdot \text{m}^{-2}$) with respect to the Ta content change. However, our estimation of $J_c$ based on Eqn. (3) clearly reflects the small variations of measured values among different samples, indicating that the SOT switching observed here is indeed better explained by a domain nucleation / domain wall propagation scenario.

In contrast to the Cu-Ta alloy case, in which the buffer layer resistivity is lower than that of pure Ta, we further engineered the buffer layer by nitrogen-doping to see the effects on resistivity enhancement. TaN buffer layer was prepared by reactive sputtering, i.e., flowing both Ar and $N_2$ gas into the sputter chamber during Ta deposition [39,40]. By adjusting the $N_2$ flow rate from 0 to 3 sccm (low doping regime) with the Ar flow rate fixed at 30 sccm, we observed a clear quasi-linear resistivity enhancement of the deposited buffer layer, as shown in Fig. 5(a). Note that beyond $N_2$ flow rate of 3 sccm (high doping regime), the resistivity of TaN has become $\geq 3000\,\mu\Omega\text{-cm}$, much greater than the resistivity of effective FM layer such that most of the applied charge current will be flowing in the FM layer instead of the spin-Hall buffer layer. The same characterization process was performed on TaN/FM/MgO Hall-bar devices and the DL-SOT efficiency $|\xi_{DL}|$ as a function of buffer layer resistivity are summarized in Fig. 5(b), together with the Cu-Ta alloy results. The quasi-linear trends in both $\rho_{CuTa} < \rho_{Ta}$ and $\rho_{TaN} > \rho_{Ta}$ regimes suggest that: (1) The DL-SOT in these heterostructures predominantly originates from the intrinsic SHE of the buffer layers; (2) The DL-SOT efficiency of a Ta-related heterostructure can be engineered by buffer layer resistivity tuning. The maximum DL-SOT efficiency found in TaN-based samples is $|\xi_{DL}| = 0.152 \pm 0.006$, almost two times the pure Ta case ($|\xi_{DL}| = 0.087 \pm 0.004$). Also note that the spin Hall conductivity in the TaN regime (slope of the DL-SOT efficiency to resistivity plot, $|\sigma_{SH}| \approx 1.84 \times 10^3 [\hbar/2e] \, \Omega^{-1} \cdot m^{-1}$) is much smaller than that in the Cu-Ta regime. This suggests that although doping can enhance $|\xi_{DL}|$, it could also change the intrinsic property of buffer layer and results in a lower $|\sigma_{SH}|$.

In summary, we report a systematic study on the correlation between DL-SOT efficiency and the buffer layer resistivity through hysteresis loop shift measurements. By varying the resistivity of $Cu_{100-x}Ta_x$ buffer layer, the DL-SOT efficiency in the Ta-rich regime is found to be linearly tunable from $|\xi_{DL}| = 0.039 \pm 0.004$ ($Cu_{84}Ta_{16}$) to $|\xi_{DL}| = 0.087 \pm 0.004$ (pure Ta). The result also suggests an intrinsic SHE mechanism in the Ta-rich buffer layer with a spin Hall conductivity

$|\sigma_{SH}| \approx 2.02 \times 10^4 [\hbar/2e] \, \Omega^{-1} \cdot m^{-1}$ (lower bound). With further systematic studies on TaN buffer layer samples, we demonstrate that the DL-SOT efficiency can also be linearly tuned up to $|\xi_{DL}| = 0.152 \pm 0.006$ by introducing $N_2$ into Ta in the low doping regime. Our work therefore provides valuable information on the engineering of magnetic heterostructure, especially the spin-Hall buffer layer, to achieve more efficient current-induced SOT switching.

**Acknowledgements**

This work was supported by the Ministry of Science and Technology of Taiwan under Grant No. MOST 105-2112-M-002-007-MY3. C. F. Pai thank Dr. Kohei Ueda for fruitful discussions.

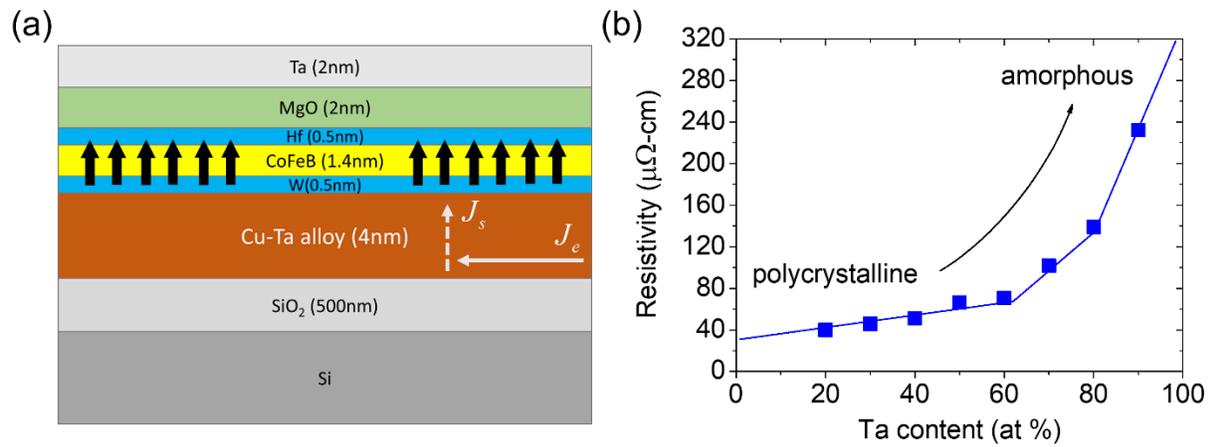

Figure 1. (a) Schematic illustration of the Cu-Ta alloy-based magnetic heterostructure. The black arrows represent the magnetic moment with desirable perpendicular magnetic anisotropy (PMA). $J_e$ and $J_s$ represent the vectors of longitudinal charge current and the transverse spin current, respectively (b) Resistivity of the co-sputtered $Cu_{100-x}Ta_x$ layer as a function of Ta content (atomic percentage).

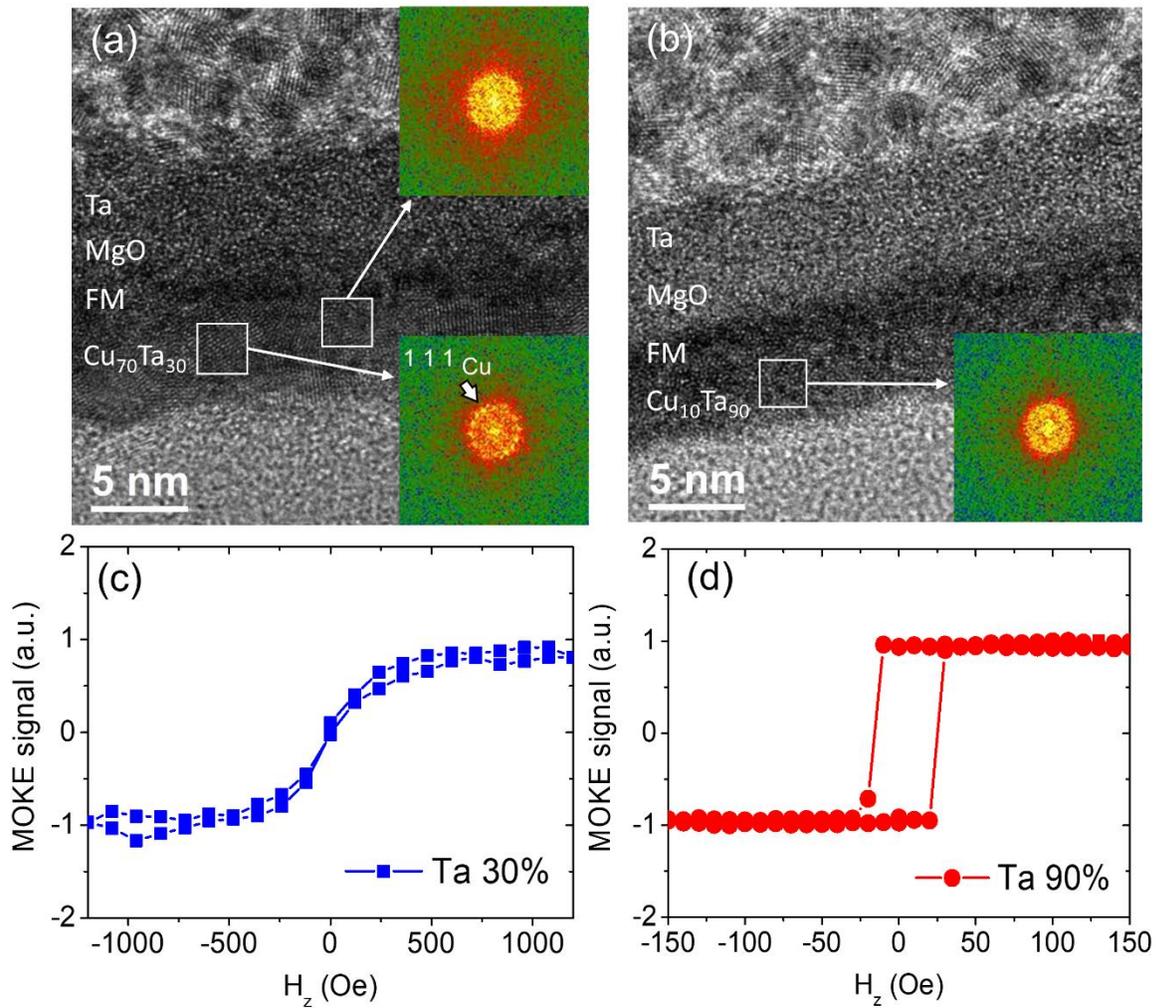

Figure 2. Cross section HR-TEM imaging results from (a) $Cu_{70}Ta_{30}$-based (Cu-rich) and (b) $Cu_{10}Ta_{90}$-based (Ta-rich) magnetic heterostructures. The subpanels show the diffractograms computed by reduced fast Fourier transformation (FFT) from the regions of interests. FM represents the effective ferromagnetic layer. Out-of-plane hysteresis loops of (c) $Cu_{70}Ta_{30}$-based (Cu-rich) and (d) $Cu_{10}Ta_{90}$-based (Ta-rich) magnetic heterostructures.

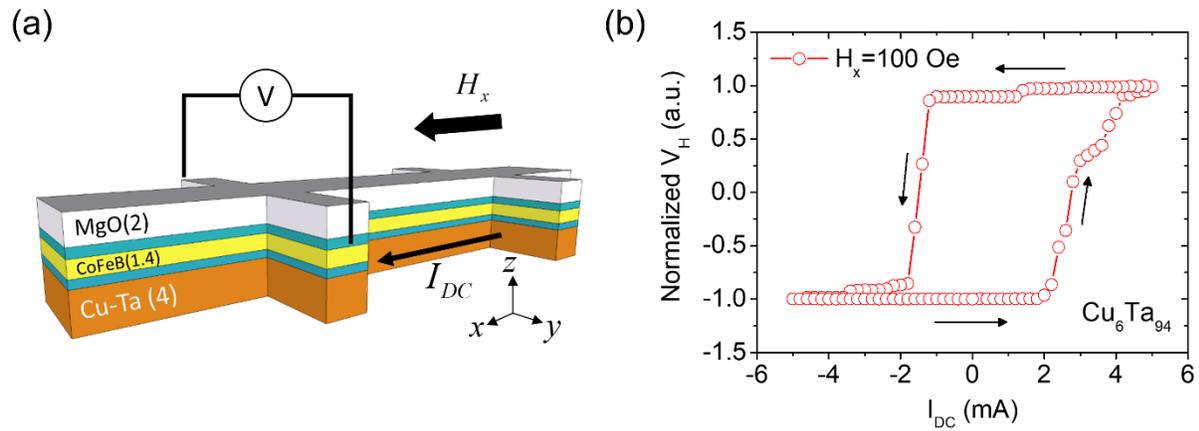

Figure 3. (a) Schematic illustration of a $Cu_{100-x}Ta_x$-based ($84 \leq x \leq 100$) Hall-bar device and the application of in-plane field $H_x$ as well as charge current $I_{DC}$ for current-induced SOT switching measurements. (b) A representative current-induced SOT switching curve of a $Cu_6Ta_{94}(4)/W(0.5)/CoFeB(1.4)/Hf(0.5)/MgO(2)$ Hall-bar device under in-plane bias field $H_x = 100\,\text{Oe}$. The black arrows indicate applied current scan directions.

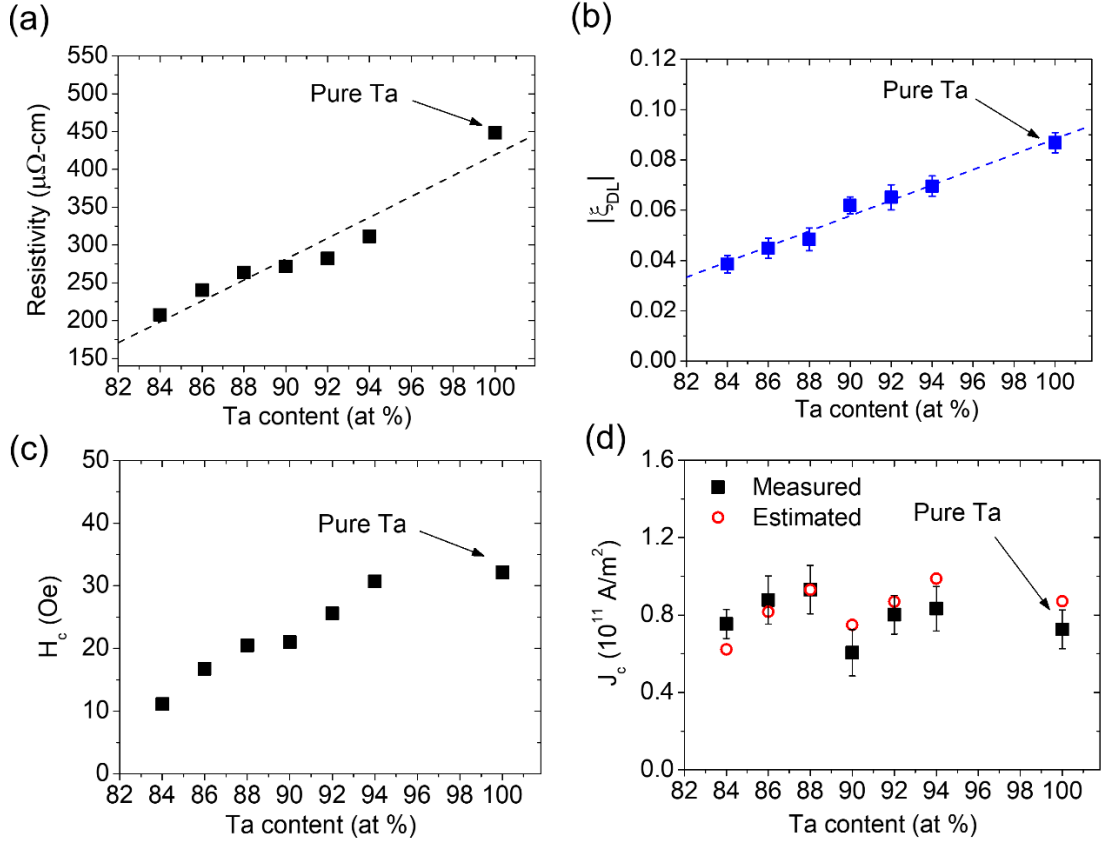

Figure 4. (a) $Cu_{100-x}Ta_x$ buffer layer resistivity, (b) the DL-SOT efficiency $|\xi_{DL}|$ of $Cu_{100-x}Ta_x$-based magnetic heterostructures, (c) coercive field $H_c$ of the magnetic layer, and (d) critical current density $J_c$ for current-induced SOT switching as functions of Ta content in the Ta-rich regime ($x \geq 84$). The dashed lines in (a) and (b) are linear fits to experimental data. The estimated results of $J_c$ (open red circles) in panel (d) are obtained from the experimental data in panels (b) and (c) through Eqn. (1).

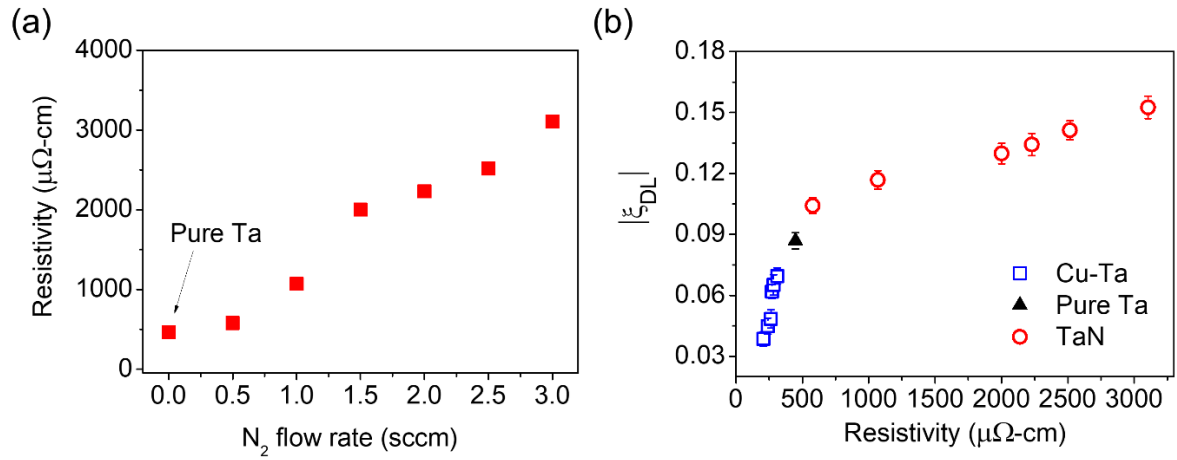

Figure 5. (a) TaN buffer layer resistivity as a function of $N_2$ flow rate during deposition. (b) A summary of characterized DL-SOT efficiency $|\xi_{DL}|$ as a function of buffer layer resistivity, for both alloying (Cu-Ta) and nitridation (TaN) samples.